\documentclass[onecolumn,showpacs,aps,prc,nofootinbib,floatfix]{revtex4}

\usepackage{amsmath}
\usepackage{bm}
\usepackage{graphicx}

\usepackage{epsfig}
\newcommand{\beq}{\begin{equation}}
\newcommand{\eeq}{\end{equation}}
\newcommand{\bea}{\vspace{0.25cm}\begin{eqnarray}}
\newcommand{\eea}{\end{eqnarray}}

\newcommand{\ro}{\mbox{{\boldmath
$\rho$}}}

\newcommand{\nb}{\mbox{{\bf
n}}}

\newcommand{\rb}{\mbox{{\bf
r}}}

\def\lsim{\mathrel{\rlap{\lower4pt\hbox{\hskip1pt$\sim$}}
    \raise1pt\hbox{$<$}}}         
\def\gsim{\mathrel{\rlap{\lower4pt\hbox{\hskip1pt$\sim$}}
    \raise1pt\hbox{$>$}}}         

\newcommand{\landau}{L.D.~Landau Institute for Theoretical Physics,
        GSP-1, 117940, Kosygina Str. 2, 117334 Moscow, Russia}

\begin{document}


\title{
  Influence of collective nuclear vibrations on initial state
  eccentricities in Pb+Pb collisions
}
\date{\today}

\author{B.G.~Zakharov}\affiliation{\landau}

\begin{abstract}
  We study within the Monte Carlo Glauber model
  the influence of collective quantum effects in
  the Pb nucleus on
  the azimuthal anisotropy coefficients $\epsilon_{2,3}$
  in Pb+Pb collisions at the LHC energies. To account for the quantum
  effects, we modify the sampling of the nucleon positions by applying
  suitable filters that guarantee that the colliding nuclei
  have the mean squared quadrupole and octupole moments consistent
  with the ones extracted from the experimental
  quadrupole and octupole strength functions for the Pb nucleus
  with the help of the energy weighted sum rule.
  Our Monte Carlo Glauber model with the modified sampling of the nucleon
  positions leads to $\epsilon_2\{2\}/\epsilon_3\{2\}\approx 0.8$ at centrality
  $\lsim 1$\%,
which allows to resolve the $v_2$-to-$v_3$ puzzle.

\end{abstract}
%

\maketitle

\section{Introduction}
        It is now believed that hadron production in heavy ion collisions
at the RHIC and LHC energies goes through quark-gluon plasma (QGP) stage.
Hydrodynamic analyses of experimental data from RHIC and LHC data
show that the QGP is formed at the proper time $\tau_0\sim 0.5-1$ fm
\cite{hydro1,hydro2,hydro3} after interaction of the Lorentz-contracted
nuclei. 
The QGP fireball, formed  between
the nuclear disks receding from each other, inherits
approximately the shape of the overlap region of the colliding nuclei.
For non-central $AA$ collisions the overlap region has an almond shape.
This can lead to a significant anisotropy in the transverse QGP expansion
at later times, and  eventually to azimuthal asymmetry of
particle spectra \cite{Olli1}. 
In the presence of fluctuations of the initial QGP density, the azimuthal
asymmetry can appear even for central collisions.
The azimuthal dependence of hadron spectra is characterized
by the flow coefficients $v_n$ in the Fourier expansion
\beq
\frac{dN}{d\phi }=\frac{N}{2\pi}
  \left\{1+{\sum_{n=1}^\infty}2v_n\mathrm{cos}\left[n\left(\phi-\Psi_n\right)
  \right]\right\},
\label{eq:10}
\eeq
where $N$ is the hadron multiplicity in a certain $p_T$ and rapidity region,
$\Psi_n$ are the event reaction plane angles.
In hydrodynamic models
with smooth initial conditions, in the Fourier
series (\ref{eq:10}) at midrapidity ($y=0$) only the terms with $n=2k$
survive (if the hadronization occurs without fluctuations). In this
approximation, for $AA$ collisions with zero impact parameter
the coefficients $v_{2k}$ should vanish due to the azimuthal symmetry.
Hydrodynamic calculations show that for heavy ion collisions
at small centralities  in each event the flow
coefficients $v_n$
with $n\le 3$ are proportional, to good accuracy, to the anisotropy
coefficients
$\epsilon_n$ for the initial entropy distribution
\cite{vn_Ollit,Niemi,vn1}
\beq
v_n\approx k_n\epsilon_n\,.
\label{eq:20}
\eeq
The coefficients $\epsilon_n$ are defined as
\cite{Teaney_en,Ollitraut_en}
\beq
\epsilon_n=\frac{\left|\int d\ro \rho^n e^{in\phi}\rho_s(\ro)\right|}
{\int d\ro \rho^n\rho_s(\ro)}\,,
\label{eq:30}
\eeq
where $\rho_s(\ro)$ is the initial fireball entropy density,
and it is assumed that $\ro$ is calculated in the 
transverse c.m. frame, i.e., $\int d\ro \ro\rho_s(\ro)=0$.

Hydrodynamic calculations require the initial entropy density,
which presently cannot be specified from first principles.
There are currently several models in use for evaluation of the initial entropy
distribution in heavy ion collisions.
The most widely used and simple phenomenological method to generate the
initial entropy distribution
  is the Monte Carlo (MC) wounded nucleon Glauber model \cite{MCGL,GLISS2},
  in which the entropy density is expressed via a linear combination
  of the number of the participating nucleons and of the binary collisions.
In the MC Glauber model event-by-event fluctuations of the entropy density
is a combined effect of fluctuations of the nucleon positions
in the colliding nuclei and fluctuations of the entropy production
for a given geometry of the nucleon positions.
The MC Glauber model has been quite successful in description
within hydrodynamic models of
experimental data on the flow
coefficients in $AA$ collisions obtained at RHIC and the LHC.
Hydrodynamic simulations with the Glauber model initial conditions
demonstrated that the QGP produced at RHIC and the LHC
has a very low shear viscosity to entropy density ratio, which is of the
order of the lower quantum bound $1/4\pi$ \cite{etas1,etas2}.
Another more recent phenomenological MC scheme for the entropy
production in $AA$ collisions, that was successfully used in hydrodynamic
analyses,
is the TRENTO model \cite{trento}. In the TRENTO model, similarly to the
MC Glauber scheme, the entropy density fluctuations originate
from fluctuations of the nucleon positions
in the colliding nuclei and fluctuations of the entropy production
for a given geometry of the positions of the participating nucleons.
This differs from the QCD-inspired IP-Glasma \cite{IP-GLASMA} and MAGMA
\cite{MAGMA} models, in which  the entropy density fluctuations
come only from fluctuations of the nucleon positions.

Although hydrodynamic models can reproduce
a vast body of data on heavy ion collisions from RHIC and the LHC,
in recent years it was found that they
have a tension with description  of the ratio between $v_2\{2\}$
and $v_3\{2\}$ in ultra-central ($c\to 0$) Pb+Pb collisions
at the LHC energies (so-called $v_2$-to-$v_3$ puzzle).
Measurements of the flow coefficients
in ultra-central $2.76$ \cite{CMS_v23} and $5.02$ \cite{ALICE_v23}
TeV Pb+Pb collisions  show that $v_2\{2\}$ and $v_3\{2\}$ are close to
each other. 
This is in disagreement with hydrodynamic calculations with the MC Glauber
and TRENTO
model initial conditions that give $v_2\{2\}/v_3\{2\}\sim 1.25-1.4$
\cite{v23_Heinz,v23_Luzum}.
This prediction is
mainly due to the fact that
for the elliptic flow the coefficient $k_2$,
in the linear response relation (\ref{eq:20}),
is bigger than the coefficient
$k_3$ for the triangular flow (e.g. for ideal hydrodynamics
the calculations performed in \cite{v23_Heinz} give $k_2/k_3\sim 1.35$
for  $2.76$ TeV  Pb+Pb collisions in $0-0.2$\% centrality bin
for $0.3<p_T<3$ GeV, and the ratio $k_2/k_3$ increases with the QGP shear
viscosity \cite{Niemi,v23_Heinz}).
Simulations of the initial entropy
distribution at zero impact parameter
within the MC Glauber and TRENTO models
give ellipticity $\epsilon_2\{2\}$ and
triangularity $\epsilon_3\{2\}$ that are close to each other
(here, as usual, $\epsilon_n\{2\}=\sqrt{\langle \epsilon_n^2\rangle}$
is the root mean squared (RMS) eccentricity).
Therefore, for $k_2/k_3>1$, the linear response relation (\ref{eq:20}) leads to $v_2\{2\}/v_3\{2\}>1$.
The problem with reproducing the experimental ratio $v_2\{2\}/v_3\{2\}$
in ultra-central Pb+Pb collisions is clearly a serious challenge
for the hydrodynamic paradigm of heavy ion collisions, because
prediction that $k_2>k_3$ seems to be quite reliable.

In recent years, there have been several attempts to explain why
in ultra-central Pb+Pb collisions
$v_2\{2\}/v_3\{2\}>1$.
In Ref.~\cite{v23_Heinz} the flow coefficients in ultra-central Pb+Pb
collisions have been addressed using
the MC Glauber and the MC-KLN \cite{KLN1,KLN2} initial conditions. It was found that
both the MC Glauber and KLN models fail to reproduce the ratio $v_2\{2\}/v_3\{2\}$.
The authors of \cite{v23_Heinz} concluded that the observed ratio
$v_2\{2\}/v_3\{2\}\approx 1$ in ultra-central Pb+Pb collisions requires
$\epsilon_2\{2\}/\epsilon_3\{2\}\sim 0.5-0.7$, which is inconsistent
both with predictions of the MC Glauber and MC-KLN models.
In Ref.~\cite{v23_Gale} the effects of bulk viscous pressure
on flow coefficients in ultra-central Pb+Pb collisions
 have been studied. It was shown
 that for the IP-Glasma initial conditions the inclusion of bulk viscosity
 can somewhat reduce $v_2\{2\}/v_3\{2\}$ ratio. Although the effect
is not strong enough to reproduce the experimental $v_2\{2\}/v_3\{2\}$ well.
In Ref.~\cite{v23_EoS} the effect of the QCD equation of state
on $v_2\{2\}/v_3\{2\}$ has been investigated for the TRENTO model initial conditions.
The authors have found that in ultra-central Pb+Pb collisions
$v_2\{2\}/v_3\{2\}\gsim 1.2$, and concluded that the variation of the
equation of state
does not allow to solve the $v_2$-to-$v_3$ puzzle.

In Refs.~\cite{v23_Heinz,v23_Gale,v23_EoS} it was assumed that
the $^{208}$Pb nucleus is spherical. 
A scenario with an octupole (pear shape) deformation of the $^{208}$Pb nucleus
has been addressed in \cite{v23_Luzum} for the TRENTO initial conditions.
This scenario seems to be appealing in the context of the $v_2$-to-$v_3$
puzzle, since for a given ratio $k_2/k_3$,
$v_2\{2\}/v_3\{2\}\propto \epsilon_2\{2\}/\epsilon_3\{2\}$. And
one can expect that the pear deformation
of the $^{208}$Pb nucleus should increase somewhat $\epsilon_3\{2\}$ (without
a significant modification of $\epsilon_2\{2\}$), and consequently
should reduce $v_2\{2\}/v_3\{2\}$.
A pear shape of the $^{208}$Pb nucleus was supported by the results of
Ref.~\cite{Bertsch} where,
within the generator-coordinate
extension of the Hartree-Fock-Bogoliubov method, the authors obtained the octupole
deformation parameter $\beta_3\sim 0.0375$ for the ground state.
However, there the value $\beta_3=0$ has been found within  the ordinary
Hartree-Fock-Bogoliubov method. More recently, in Ref.~\cite{Afan}
the value $\beta_3=0$ for the $^{208}$Pb nucleus ground state
has also been obtained within the covariant density functional theory.
The results of \cite{v23_Luzum} show that, for reasonable values
of $\beta_3$, the scenario with the octupole deformation
of the $^{208}$Pb does not lead to a significant improvement in description
of the ratio $v_2\{2\}/v_3\{2\}$ in ultra-central Pb+Pb collisions.

In the studies
\cite{v23_Heinz,v23_Gale,v23_EoS,v23_Luzum},
the initial conditions have been generated using the MC sampling of the nucleon
positions with the Woods-Saxon (WS) nuclear distribution.
In fact, presently this method is the standard approach for MC calculations
of the initial conditions in heavy ion collisions.
One of the apparent shortcomings of the MC WS sampling of the nucleon
positions is that
this approach completely ignores the collective dynamical effects
for the long range fluctuations of the nucleon positions
(which are especially important for calculations of the azimuthal
anisotropy coefficients $\epsilon_n$).
Indeed, it is well known that the long range nuclear density
fluctuations have a collective nature \cite{BM,Greiner}.
The collective effects manifest themselves in the presence
of giant resonances/vibrations, which correspond to coherent oscillations
of the nucleons \cite{BM,Greiner}
(for more recent reviews see \cite{Speth,Roca}).
Since the long range collective effects are ignored in the MC WS sampling
of the nucleon positions,
there is no guarantee that this approach can mimic the true
long range fluctuations of the nuclear density.
One can expect that, in the context of the anisotropy coefficients
$\epsilon_2$ and $\epsilon_3$ for ultra-central Pb+Pb collisions,
the most crucial giant vibration modes are the quadrupole and octupole ones.
In our previous work \cite{Z1}, we have investigated the possible effect
of the isosinglet quadrupole giant vibration mode on the $\epsilon_2\{2\}/\epsilon_3\{2\}$ ratio.
The analysis \cite{Z1} was motivated by the fact, established in
\cite{Z_e2}, that the MC WS sampling of the nucleon positions
leads to a significant overprediction of the mean squared quadrupole
moment of the $^{208}$Pb nucleus as compared to that
obtained via the experimental parameters of the
isosinglet giant quadrupole resonance (ISGQR).
The quantum calculation with the  help of the energy weighted sum
rule (EWSR) for the quadrupole strength function (for a review, see \cite{EWSR})
gives the mean squared quadrupole
moment that is smaller
than the one calculated with the WS nuclear density by a factor of $\sim 2.2$
\cite{Z1} (after correcting an error made in \cite{Z_e2}).
This means that prolate and oblate elliptic fluctuations of the $^{208}$Pb
nucleus are significantly weaker than predicted by the MC WS sampling
of the nucleon positions.
For this reason, one can expect that the true many-body nuclear density
should give a smaller ellipticity $\epsilon_2$ than the MC simulation
with the standard WS nuclear density.
To study quantitatively this
effect, in \cite{Z1} we have performed the MC Glauber model calculations
of the anisotropy coefficients $\epsilon_{2,3}$ for central Pb+Pb collisions
using a modified method of the MC sampling of the nucleon positions that
guarantees that the averaged over all collisions squared quadrupole
moments of the colliding
nuclei coincide with the mean squared quadrupole
moment of the $^{208}$Pb nucleus obtained using the EWSR.
The results of \cite{Z1} show that the modified MC sampling with filtering
the nucleon positions by the value of the quadrupole moment
leads to a noticeable reduction
of the ellipticity $\epsilon_2$. It was found that the quadrupole moment
filtering practically does not change the prediction for the
triangularity $\epsilon_3$.
We have obtained that the MC Glauber model with 
the quadrupole moment filtering of the nucleon positions 
gives
$\epsilon_2\{2\}/\epsilon_3\{2\}\approx 0.8$ for 2.76 and 5.02 TeV
central Pb+Pb collisions. 
Then, adopting the hydrodynamic linear response coefficients $k_{2,3}$
from \cite{Olli_k23,v23_Heinz,Olli_Xe,v23_Luzum}),
we obtained $v_2\{2\}/v_3\{2\}\approx 0.96-1.12$
which agrees reasonably with the data from ALICE \cite{ALICE_v23}. 

One weakness of the analysis \cite{Z1} is that the central
Pb+Pb collisions are treated as collisions at zero impact parameter,
i.e., calculations of Ref. \cite{Z1} correspond to
$b$-centrality, in the terminology of Ref.~\cite{Ollit_cent},  defined
in terms of the impact parameter $b$
($c=\pi b^2/\sigma_{in}^{AA}$ \cite{centrality}). 
However, experimentally, the centrality of a collision
is usually estimated  through charged particle multiplicity $N_{ch}$
in a certain kinematic region. This $n$-centrality is defined as
\cite{centrality,Ollit_cent}
\beq
c(N_{ch})=\sum_{N=N_{ch}}^{\infty}P(N)\,,
\label{eq:40}
\eeq
where $P(N)$ is the probability for observing the multiplicity $N$.
Due to multiplicity fluctuations (at a given impact parameter),
there is some mismatch between
the $b$- and $c$-centralities \cite{centrality,Ollit_cent}. For this reason,
one can reasonably worry about the effect of this mismatch
on the results of Ref.~\cite{Z1} where the effect of $n$-centrality smearing
at a given $b$-centrality has been ignored.
Therefore, it is highly desirable to extend the calculations of \cite{Z1}
to the case of the $n$-centrality.
This is our main purpose in the present paper.
Also, we extend the analysis of \cite{Z1} to the case of the octupole
fluctuations. 
Investigation of the role of the filtering the nucleon positions
by the octupole moment
in the MC simulations of Pb+Pb collisions is interesting because
the collective pear shape fluctuations may potentially affect the triangularity
$\epsilon_3$ of the fireball.
From the available experimental data
one can conclude that for the octupole shape fluctuations the mean
squared octupole  moment  of the $^{208}$Pb nucleus may be
somewhat larger than the one obtained from the MC WS calculations (see appendix).
  The latter possibility seems to be very interesting in the context
  of the $v_2$-to-$v_3$ puzzle, because it should lead to an increase of
  $\epsilon_3$ (similarly to the case with the pear shape deformation of the
  ground state \cite{v23_Luzum}) and, consequently, to a smaller value of the ratio
  $\epsilon_2\{2\}/\epsilon_3\{2\}$.
  Note that contrary to the analysis of Ref.~\cite{Z1}, in the present
  work we perform calculations for the whole range of centrality.
  As in \cite{Z1}, we use the MC Glauber model developed in \cite{Z_gl1,Z_gl2},
  which allows to account for the presence of the meson-baryon component
  in the nucleon light-cone wave function.

  The plan of the paper is as follows. In Sec.~2 we 
  discuss the theoretical framework.
In Sec.~3 we present our numerical results.
We give conclusions in Sec.~4.
In appendix we discuss calculations of the mean squared quadrupole
and octupole moments
of the $^{208}$Pb nucleus using the EWSR.

\section{Theoretical framework}  

In the present analysis, to generate the initial entropy density
we use the MC Glauber  approach developed in
\cite{Z_gl1,Z_gl2}.
This MC Glauber model allows to perform calculations of the entropy production
in the standard way,
when each nucleon is treated as a one-body state, and 
accounting for the presence of the meson cloud in the nucleon,
when the physical nucleon light-cone function includes
the bare nucleon and meson-baryon Fock states.
The results of our previous analyses \cite{Z_gl2,Z_gl3} show that,
for both the versions, the predictions of this model for centrality dependence
of the midrapidity charged multiplicity density are in very
good agreement with experimental data for $0.2$ TeV Au+Au
collisions at RHIC, $2.76$ and $5.02$ TeV Pb+Pb, and 
$5.44$ TeV Xe+Xe collisions at the LHC.

\subsection{Outline of the MC Glauber scheme}  
In this subsection we briefly outline the algorithm used in our MC Glauber model for
the version without the meson-baryon component
(in this case, our scheme is similar to the MC Glauber generator
GLISSANDO \cite{GLISS2}).
The entropy generation
occurs through the wounded nucleons (WNs) and through the
hard binary collisions (BCs). We assume that for each pair of colliding
nucleons the cross section of the
hard binary collision is suppressed by a factor $\alpha$ \cite{KN}. 
The total entropy density in the transverse plane is written as
(we consider the central rapidity region)
\beq
\rho_s(\ro)=\sum_{i=1}^{N_{wn}} S_{wn}(\ro-\ro_i)+
\sum_{i=1}^{N_{bc}} S_{bc}(\ro-\ro'_{i})\,,
\label{eq:50}
\eeq
where the $S_{wn}$ terms corresponds to
the WN sources  and $S_{bc}$ terms
to the BC sources, $N_{wn}$ and $N_{bc}$ are the numbers of the WNs
and BCs, respectively. We write $S_{wn}$ and $S_{bc}$ as
\beq
S_{wn}(\ro)=\frac{(1-\alpha)}{2}s(\ro)\,,\,\,\,\,\,
S_{bc}(\ro)=s(\ro)\,,
\label{eq:60}
\eeq
where 
$s(\ro)$ is the source entropy distribution. We use for $s(\ro)$ a Gaussian form
\beq  
s(\ro)=s_0\exp{\left(-\ro^2/\sigma^2\right)}/\pi \sigma^2\,
\label{eq:70}
\eeq
with $s_0$ the total entropy of the source, and $\sigma$ width of the source.  
We assume that the center of each WN entropy source coincides with
the WN position, and  for each BC the center of the entropy source
is located in the middle between the colliding nucleons.

For each entropy source, we treat $s_0$ as a random variable.
We assume that the QGP expansion
is isentropic. In this approximation we can treat each entropy source as
a source of the charged multiplicity $n=a s_0$
in the unit pseudorapidity  interval $|\eta|<0.5$ with
$a\approx 7.67$ \cite{BM-entropy}.
We describe the fluctuations of $n$ by the Gamma distribution
\beq
\Gamma(n,\langle n\rangle)=
\left(\frac{n}{\langle n\rangle}\right)^{\kappa-1}
\frac{\kappa^\kappa\exp\left[-n\kappa/\langle n\rangle\right]}
{\langle n\rangle \Gamma(\kappa)}\,
\label{eq:80}
\eeq
with the parameters $\langle n\rangle$ and $\kappa$ adjusted to
fit the experimental mean charged multiplicity 
and its variance in the unit pseudorapidity window $|\eta|<0.5$
for $pp$ collisions.

As in the analyses \cite{Z_gl2,Z_gl3}, in the version with the meson-baryon
component of the nucleon,
for the total weight of the $MB$ states in the physical nucleon we
take $40$\%
that allows one to describe the DIS data on the violation of the
Gottfried sum rule \cite{ST}.
In the sense of the entropy sources, calculation of the initial entropy density in this version
is similar to that for the version without $MB$ component. However,
in this case, the entropy sources can be produced in $BB$, $MB$, and $MM$
collisions.
The results of \cite{Z_gl2,Z_gl3} show that
both the versions give similar predictions for the midrapidity charged
multiplicity density $dN_{ch}/d\eta$. However, the version with the $MB$ component
requires somewhat smaller value of the parameter $\alpha$
to fit the measured midrapidity $dN_{ch}/d\eta$.
In the present analysis we use the values $\alpha=0.14$($0.09$)
for the versions without(with) the meson-baryon component of the nucleon.
These values allow to reproduce very well
the data on the centrality dependence of $dN_{ch}/d\eta$
at $\eta=0$ for $2.76$ and $5.02$ TeV Pb+Pb collisions.
For more details on our MC Glauber scheme we refer the reader to
Refs.~\cite{Z_gl2,Z_gl3}.

\subsection{Sampling of nucleon positions}
The MC Glauber model gives the algorithm for calculation of the entropy
distribution in each $AA$ collision for given nucleon positions
in the colliding nuclei. It should be supplemented
by a prescription for the MC sampling of the nucleon positions.
Usually, in event-by-event simulations of heavy ion collisions,
the nucleon positions are generated using the uncorrelated WS distribution
(or the WS distribution with
a restriction on the minimum distance between two nucleons
\cite{Bron_NNcore,GLISS2} to model the $NN$ hard core).
However, this procedure completely
ignores the collective nature of the long range fluctuations
of the nuclear density, and can lead to an incorrect description
of the 3D fluctuations of the many-body density of the
colliding nuclei.
This can translate to incorrect predictions for fluctuations of the initial
entropy distribution in $AA$ collisions.
As already mentioned in the
introduction, from the point of view of heavy ion collisions, the most
important collective fluctuations are related to the quadrupole and
octupole vibration modes.
Their magnitude can be characterized by
the squared $L$-multipole moment (we denote it as $Q_L^2$)
for $L=2$ and $3$ defined via the spherical harmonics (see appendix).
In \cite{Z1} we have suggested a simple systematic method for calculations
of the mean squared multipole moments, $\langle Q_L^2\rangle$, for arbitrary $L$
from experimental strength functions using
the EWSR (for completeness, in appendix we outline it).
For the $L=2$ mode this method gives the mean squared quadrupole moment
of the $^{208}$ Pb nucleus
that is smaller than predicted by the MC simulation with the WS nuclear density
by the factor $r_2\approx 2.25$ (see appendix).
One can expect that the overprediction
of the $L=2$ fluctuations of the $^{208}$Pb nuclear density
may translate to an overprediction of the ellipticity $\epsilon_2$
in the MC simulation of ultra-central Pb+Pb collisions.
In Ref. \cite{Z1} we have proposed a simple method for curing this problem
by performing  the MC sampling of the nucleon positions with a suitable
$Q_2^2$-filter, which should guarantee the true value of 
$\langle Q_2^2\rangle$
for the final sample of the nucleon positions.
In Ref. \cite{Z1} we have performed calculations using two different
$Q_2^2$-filters with smooth and sharp filtering.
In the smooth version we used a $Q_2^2$-filter 
that generated a set of the nucleon positions with distribution
in $Q_2^2$
which was equal to the rescaled by the factor $r_2$ $Q_2^2$ distribution
for the WS nuclear distribution.
In the second method
we merely selected only the nucleon configurations with
$Q_2^2<Q_{2\text{max}}^2$ with $Q_{2\text{max}}^2$ adjusted to provide in the MC sampling
the correct EWSR $\langle Q_2^2\rangle$. It was found that
these two very different filters give practically identical
results for $\epsilon_{2,3}\{2\}$.

As in Ref. \cite{Z1}, in the present analysis we perform calculations
using smooth and sharp $Q_2^2$-filterings of the nucleon positions.
In the first case, we use in the MC sampling of the nucleon positions
a smooth $Q_2^2$-filter which generates the nucleon positions
with the $Q_2^2$ distribution given by
\beq
P(Q_2^2)=C\exp(-(Q_2^2/a_2)^2)P_{WS}(Q_2^2)\,,
\label{eq:90}
\eeq
where $P_{WS}$ is the $Q_2^2$ distribution for the ordinary unfiltered
MC WS sampling of the nucleon positions, $C$ is the normalization constant,
and $a_2$ is the parameter adjusted
to have $\langle Q_2^2 \rangle=\langle Q_2^2 \rangle_{WS}/r_2$.
From the point of view of numerical computations, the ansatz (\ref{eq:90})
with the Gaussian suppression factor $\exp(-(Q_2^2/a_2)^2)$
is
simpler than the method  of \cite{Z1}, with rescaling the original
WS distribution $P_{WS}(Q_2^2)$.
In the second method, as in \cite{Z1}, we use a
sharp filter with a
cutoff $Q_2^2<Q_{2\text{max}}^2$ with $Q_{2\text{max}}^2$ adjusted
to have $\langle Q_2^2 \rangle=\langle Q_2^2 \rangle_{WS}/r_2$.
As in \cite{Z1}, we have found that predictions for $\epsilon_{2,3}$
obtained for the smooth and sharp $Q_2^2$-filters are practically
indistinguishable.

In the present analysis, in addition to the effect
of the quadrupole vibrations addressed in \cite{Z1},
we also study the influence on the anisotropy coefficients
$\epsilon_{2,3}$ of the octupole ($L=3$) vibrations
of the $^{208}$Pb nucleus.
Similarly to the case of the quadrupole fluctuations of the
3D nuclear density, an inappropriate description
of the octupole 3D nuclear density fluctuations in the MC WS
sampling of the nucleon positions can lead to incorrect predictions for
the 2D initial
entropy fluctuations in Pb+Pb collisions. It is reasonable to expect
that for ultra-central Pb+Pb collisions the changes in the octupole
3D fluctuations of the nuclear density
will mostly affect the triangularity $\epsilon_3$.

Unfortunately, there are rather large uncertainties
in the experimental data on the octupole strength function
of the $^{208}$Pb nucleus (see appendix),
which translate to considerable uncertainties in the value of the
mean squared octupole moment obtained using the EWSR.
Calculations using the EWSR
and the available data on the octupole strength function of the $^{208}$Pb
nucleus,
show that the ratio of the mean squared octupole moment predicted by the WS
$^{208}$Pb nuclear density to the true one
should most likely lie within the range $0.7< r_3<0.84$
(see appendix).  
Thus, contrary to the situation with the quadrupole mode,
it is possible that
the WS nuclear density somewhat underpredicts the 3D octupole fluctuations
of the $^{208}$Pb nucleus.
To model the effect of possible enhancement of the octupole fluctuations
for the $^{208}$Pb nucleus on the initial entropy distribution, we
use, similarly to the case of the quadrupole mode, two types
of filters in the sampling of the nucleon positions.
In this first method, we use a smooth $Q_3^2$-filter, which
generates the nucleon positions with the distribution
in $Q_3^2$ given by
\beq
P(Q_3^2)=C[1-\exp(-(Q_3^2/a_3)^2)]P_{WS}(Q_3^2)\,.
\label{eq:100}
\eeq
In the second method, we use a sharp filter that selects
only the configurations with $Q_3^2>Q_{3\text{min}}^2$.
The values of $a_3$ and $Q_{2\text{max}}^2$ are adjusted
to have $\langle Q_3^2 \rangle=\langle Q_3^2 \rangle_{WS}/r_3$.
Both these prescriptions push the $\langle Q_3^2 \rangle$ to higher values.
As for the $L=2$ mode, we have found that predictions for $\epsilon_{2,3}$
obtained for the smooth and sharp $Q_3^2$-filters are practically
identical.
It is worth noting that although our $Q_{2,3}^2$-filters give significant
changes in the distributions in $Q_{2,3}^2$ for the generated
set of the nucleon positions, they give almost
zero effect on the one-nucleon density distribution (i.e., after the
$Q_{2,3}^2$-filtering we have the same WS density distribution).  

In Fig. 1a we plot the distributions in the squared $L=2$
multiple moment obtained for the ordinary MC sampling of the nucleon
positions for uncorrelated WS density
of the $^{208}$Pb nucleus without and with $Q_2^2$-filtering
(for the smooth $Q_2^2$-filter that corresponds to $r_2=2.25$).
In Fig. 1b we show similar results for
the $L=3$ mode. For this mode  we show the results for two filtered
distributions for $r_3=0.84$ and $0.7$.
We use in Fig. 1 the dimensionless variables $q_L=Q_L^2/AR_A^{2L}$,
where $R_A$ is the nucleus radius in the WS parametrization
of the $^{208}$Pb nuclear density (\ref{eq:a10}).

\begin{figure}[!h] 
\begin{center}
\includegraphics[height=5.5cm]{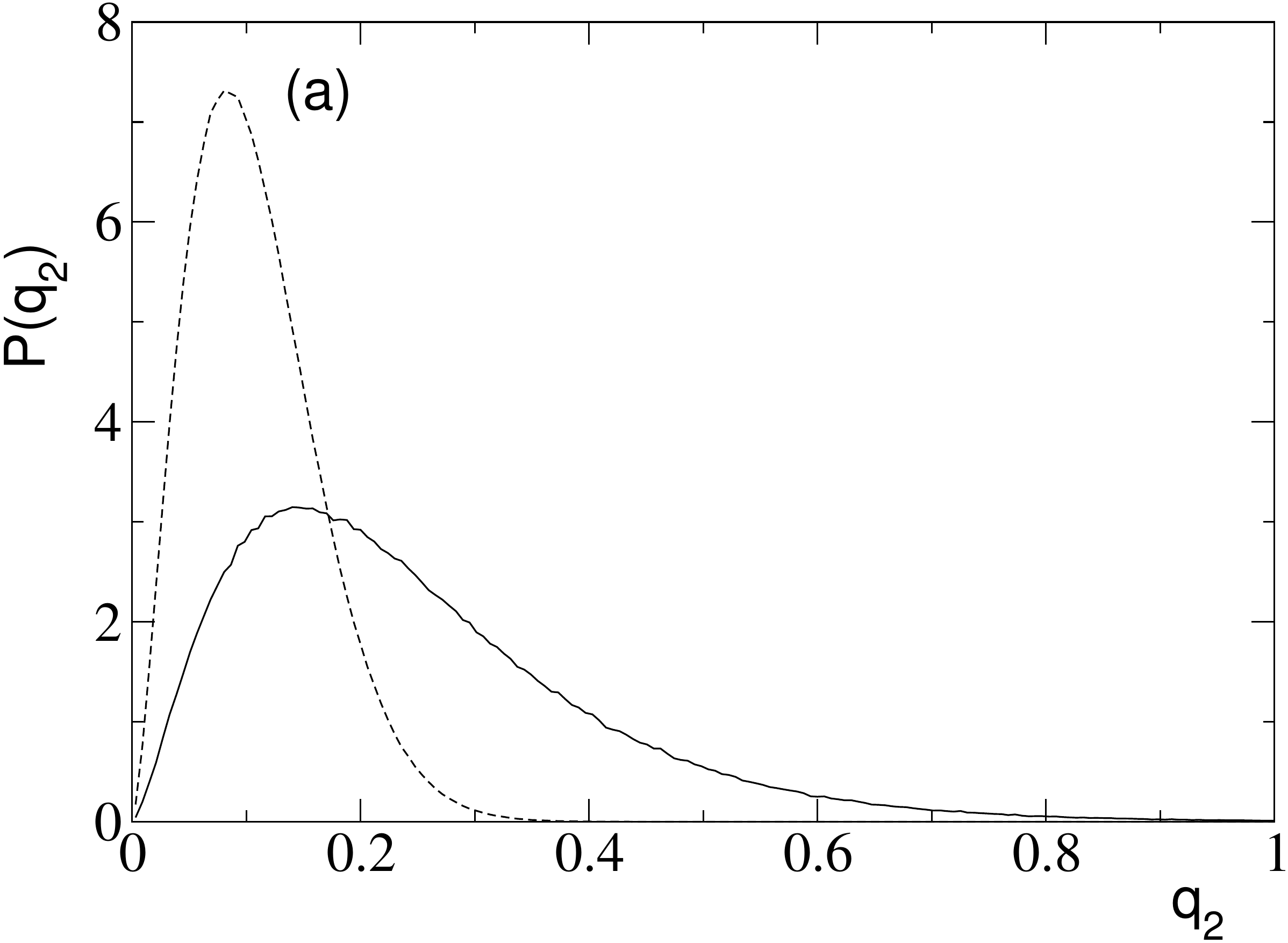}  
\includegraphics[height=5.5cm]{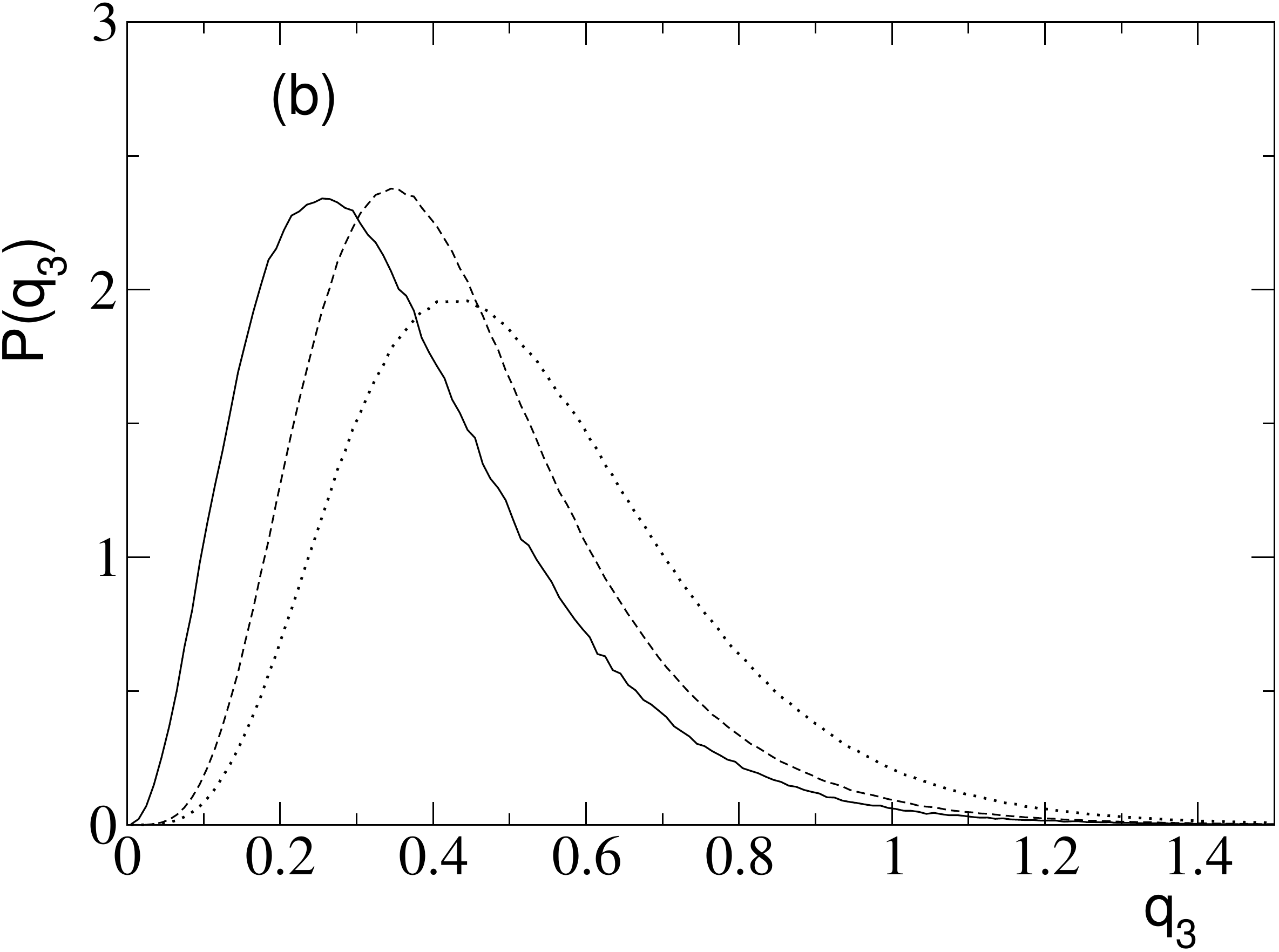}  
\end{center}
\caption[.]{ (a) The distribution in the squared quadrupole moment
  in terms of the dimensionless variable $q_2=Q_2^2/AR_A^4$
  for the $^{208}$Pb nucleus obtained using the ordinary MC sampling 
  of the nucleon positions for the WS nuclear density (solid) and
  with filtering the nucleon positions (dashed) that gives the mean squared
  quadrupole moment reduced by the factor $r_2\approx 2.25$
  (see text for explanations). (b)
 The distribution in the squared octupole moment
  in terms of the dimensionless variable $q_3=Q_3^2/AR_A^6$
  for the $^{208}$Pb nucleus obtained using the ordinary MC sampling 
  of the nucleon positions for the WS nuclear density (solid) and
  with filtering the nucleon positions that gives the mean squared
  octupole moment increased by the factor $1/r_3$
  for $r_3=0.84$ (dashed) and $0.7$ (dotted).
 }
\end{figure}

It is worth noting that our numerical calculations show that
the $Q_2^2(Q_3^2)$-filtering practically does not affect the
$Q_3^2(Q_2^2)$ distribution. This occurs because to very good accuracy
the original two dimensional distribution in $Q_{2,3}^2$,
for the MC WS sampling of the nucleon positions, can be written
in a factorized form
\beq
P_{WS}(Q_2^2,Q_3^2)\approx P_{WS}(Q_2^2)P_{WS}(Q_3^2)\,.
\label{eq:110}
\eeq
Note that, similarly to the cases when the $Q_2^2$- and $Q_3^2$-filters
are applied separately,
our numerical calculations show that for simultaneous use of 
the $Q_2^2$- and $Q_3^2$-filters predictions for $\epsilon_{2,3}$
turn out to be practically identical for the smooth and sharp filters.

We have also investigated the effect of the modification
of the distribution in the isovector dipole moment.
For the isovector dipole fluctuations, the MC sampling of the nuclear
configurations with the WS nuclear density leads to the mean squared
dipole moment that is larger by a factor of $\sim 5-6$ 
than that obtained from the parameters of the isovector dipole resonance \cite{Z_e1,Z_e2}.
The isovector giant dipole resonance corresponds to the collective
oscillations of the protons and neutrons in opposite directions
\cite{BM,Greiner}. This mode can lead to a prolate form
of the nucleon distribution (i.e. it generates some quadrupole moment),
and in principle inadequate description of this mode could affect the geometry
of the entropy distribution in Pb+Pb collisions.
However, we have found the effect of the modification
of the MC sampling of the nucleon positions for the isovector
dipole mode (in the same way as we do it for the isosinglet quadrupole
mode) on the results for $\epsilon_{2,3}$ turns out to be
practically negligible. Physically, this is due to a very small
statistical weight (among the quadrupole fluctuations)
of the fluctuations with the collective displacement
of all the protons and all the neutrons in opposite directions.
Therefore, a modification of the the distribution in the isovector
dipole moment in the MC sampling of the nucleon
positions gives almost zero effect on  $\epsilon_{2,3}$.

It is worth noting that the sampling of the nucleon positions
for the WS nuclear density leads to some overprediction
of the pure radial fluctuations, corresponding to the monopole
($L=0$) vibration mode, as 
 compared to prediction from the EWSR for the experimental monopole strength
function (see appendix). 
However, intuitively one could expect that the effect of the radial fluctuations
should be immaterial for the eccentricities $\epsilon_{2,3}$
(especially at small centralities), and the disagreement between the $L=0$
moments for the WS sampling of the nucleon positions and that obtained from
the EWSR should not be important. 
Our calculations confirm this, we have found that adding the filtering
for the $L=0$
mode practically does not affect the azimuthal coefficients $\epsilon_{2,3}$,
hence we have not used a filter for the $L=0$ mode.

Finally, we would like to emphasize that 
the fact that all our predictions for $\epsilon_{2,3}\{2\}$ for the smooth and sharp
filters are practically identical is very encouraging from the point of
view of the validity of the
strategy to mimic collective effects by simple $Q_{2,3}^2$-filterings the
nucleon positions.
Indeed, our smooth and sharp filters lead to radically different
distributions  in $Q_2^2$ and $Q_3^2$. It is clear that the many-body
densities for these filters are also radically different.
Nevertheless, we obtain
practically identical $\epsilon_{2,3}\{2\}$, if both the versions
lead to the same values of $\langle Q_2^2 \rangle$
and $\langle Q_3^2 \rangle$, and the difference in their
other characteristics (say, the difference in the values of $\langle
(Q_{2,3}^2)^2 \rangle$) has a negligible effect on $\epsilon_{2,3}\{2\}$
\footnote{The reason for this property of $\epsilon_{2,3}\{2\}$ is unclear.
It may be connected with the fact
  that in the Glauber wounded nucleon model the variance of $\epsilon_n$
  (as $\langle Q_{2,3}^2 \rangle$)
  depends only on the two-nucleon correlators for the colliding nuclei.
While $\langle(Q_{2,3}^2)^2 \rangle$ depend on the four-nucleon correlators
as well, which are not important for the variance of $\epsilon_n$ at all.}.  
This feature of the Glauber model predictions for
$\epsilon_{2,3}\{2\}$ allows us to expect
that our results for $\epsilon_{2,3}\{2\}$ should be close to those for
the true many-body density, provided that we
use the $Q_{2,3}^2$-filters guaranteeing the correct values of
$\langle Q_2^2\rangle$ and $\langle Q_3^2 \rangle$.

\section{Numerical results for $\epsilon_2\{2\}$ and $\epsilon_3\{2\}$}
\begin{figure}[!h] 
\begin{center}
\includegraphics[height=5.5cm]{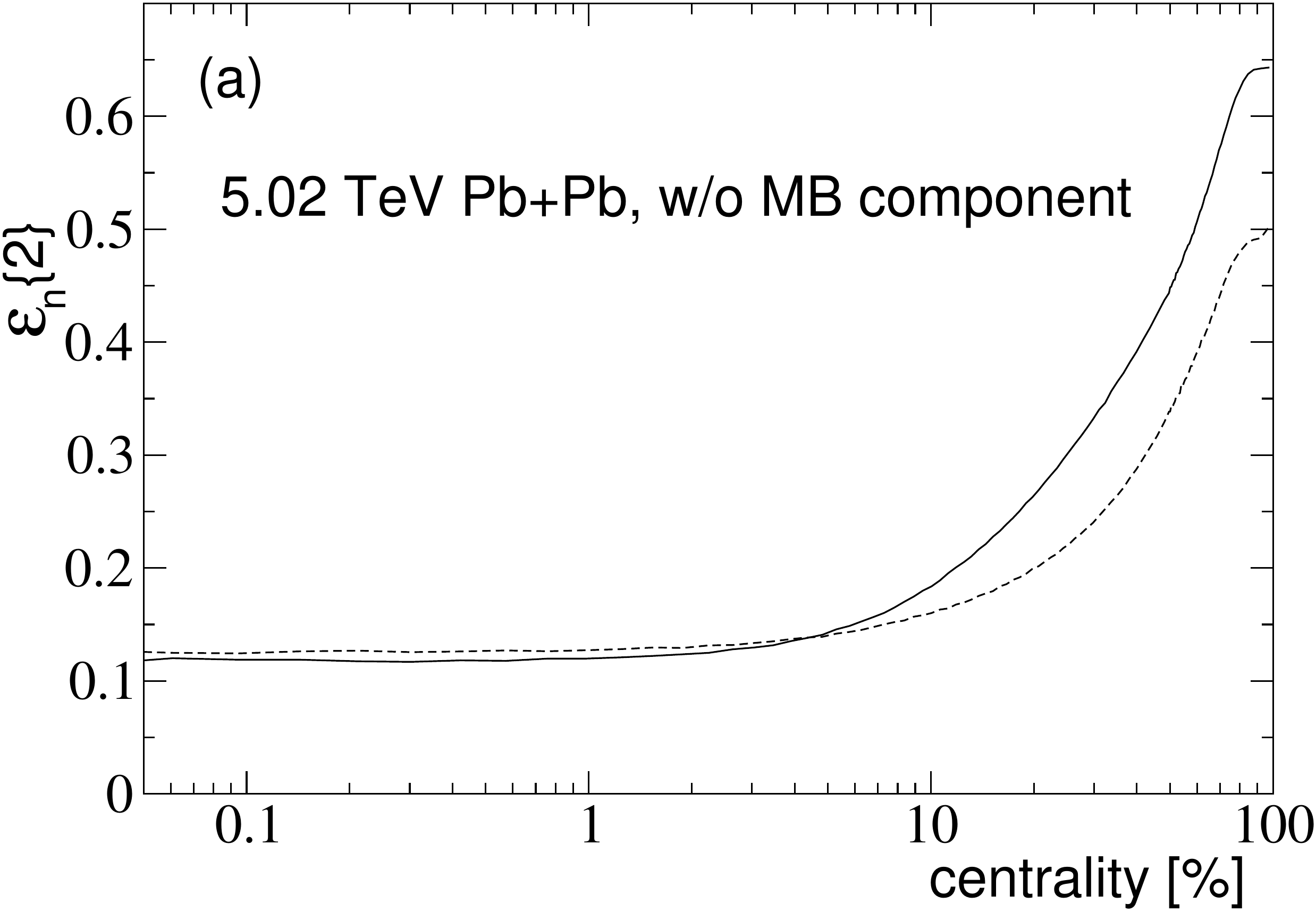}  
\includegraphics[height=5.5cm]{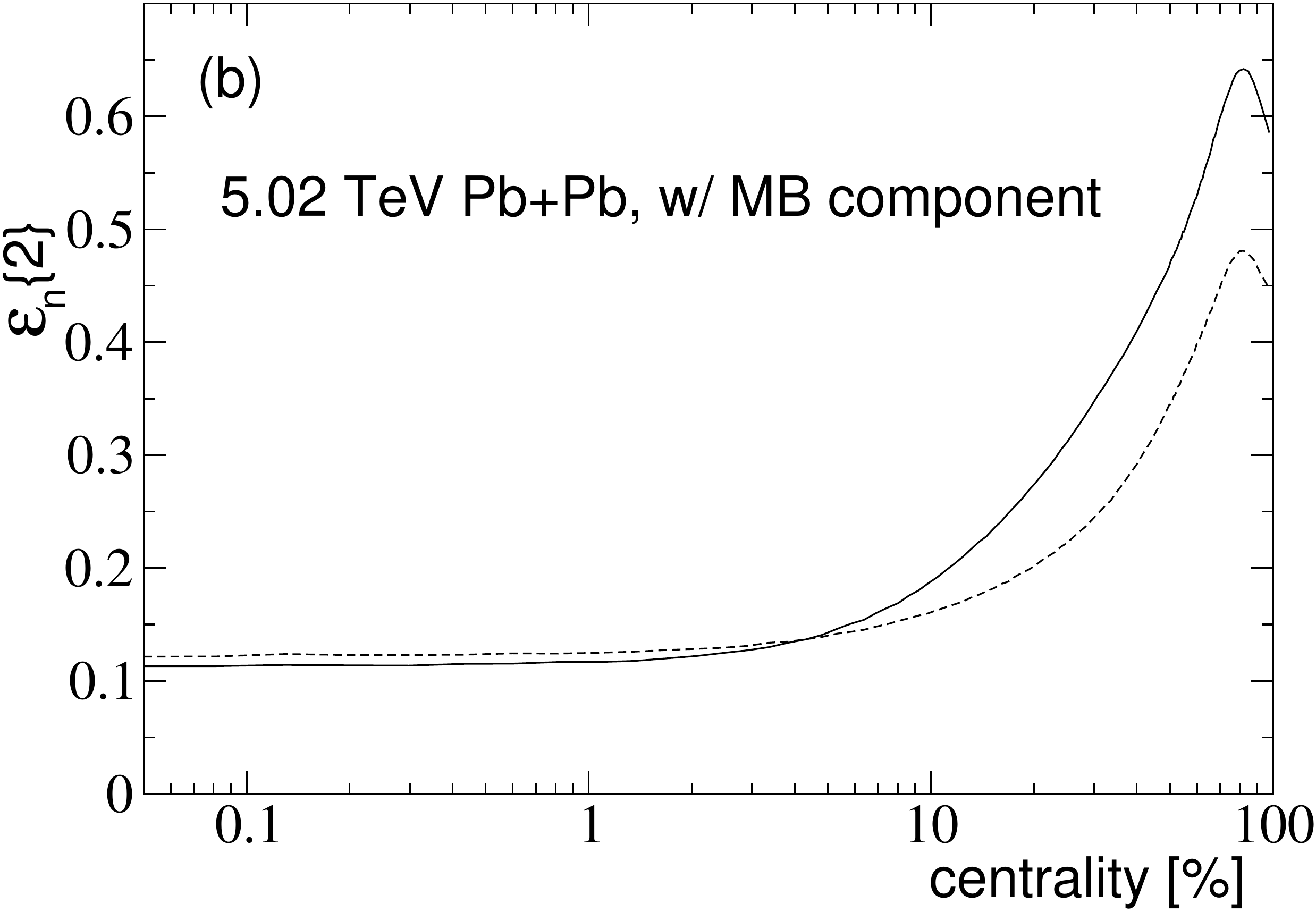}  
\end{center}
\caption[.]{The rms azimuthal coefficients $\epsilon_{2}\{2\}$ (solid)
  and $\epsilon_3\{2\}$ (dashed) vs centrality
  for 5.02 TeV Pb+Pb collisions obtained within the MC Glauber model
  without (a) and with (b) the meson-baryon component of the nucleon
  using the ordinary MC WS sampling of the nucleon positions.
 }
\end{figure}
\begin{figure}[!h] 
\begin{center}
\includegraphics[height=5.5cm]{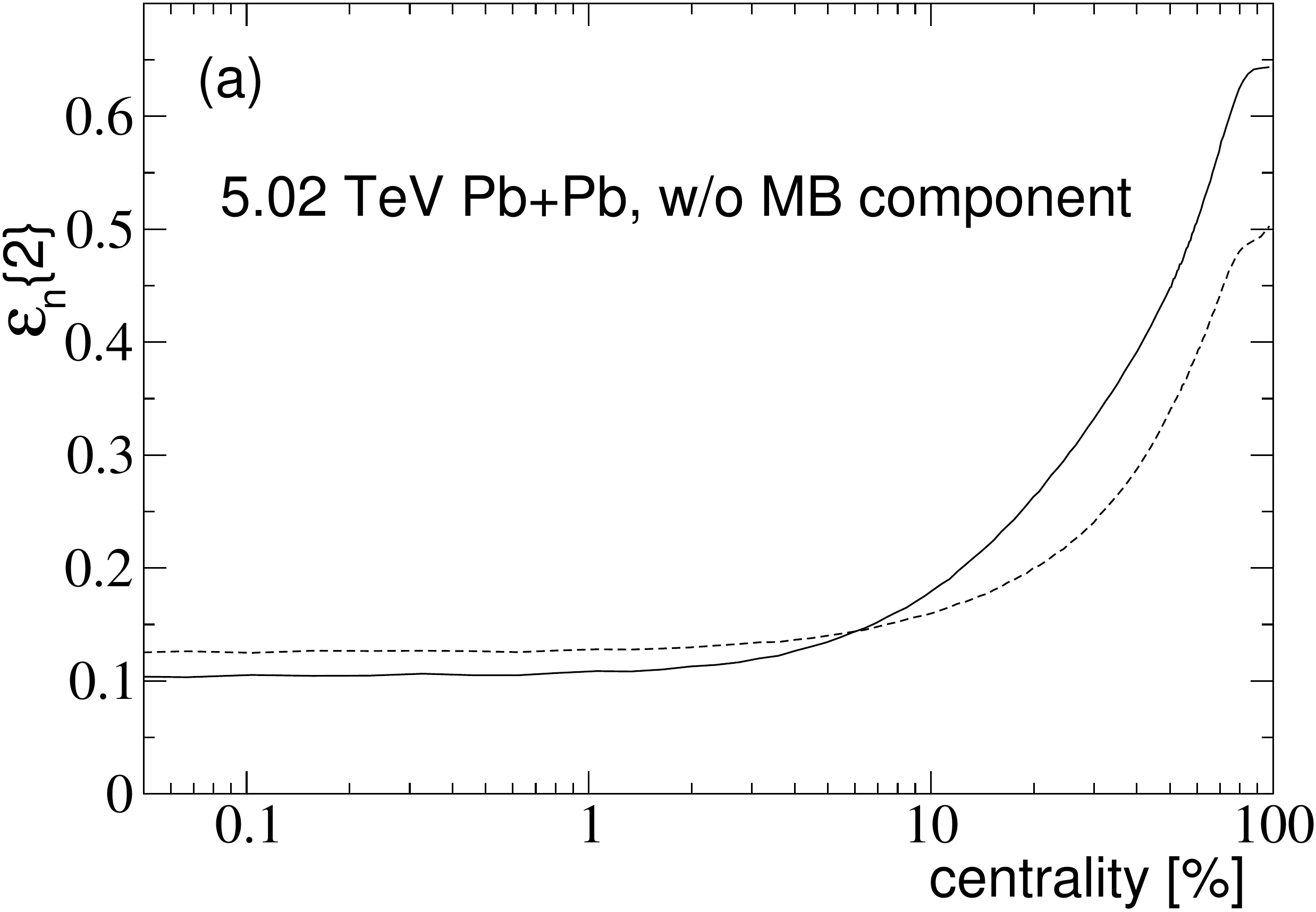}  
\includegraphics[height=5.5cm]{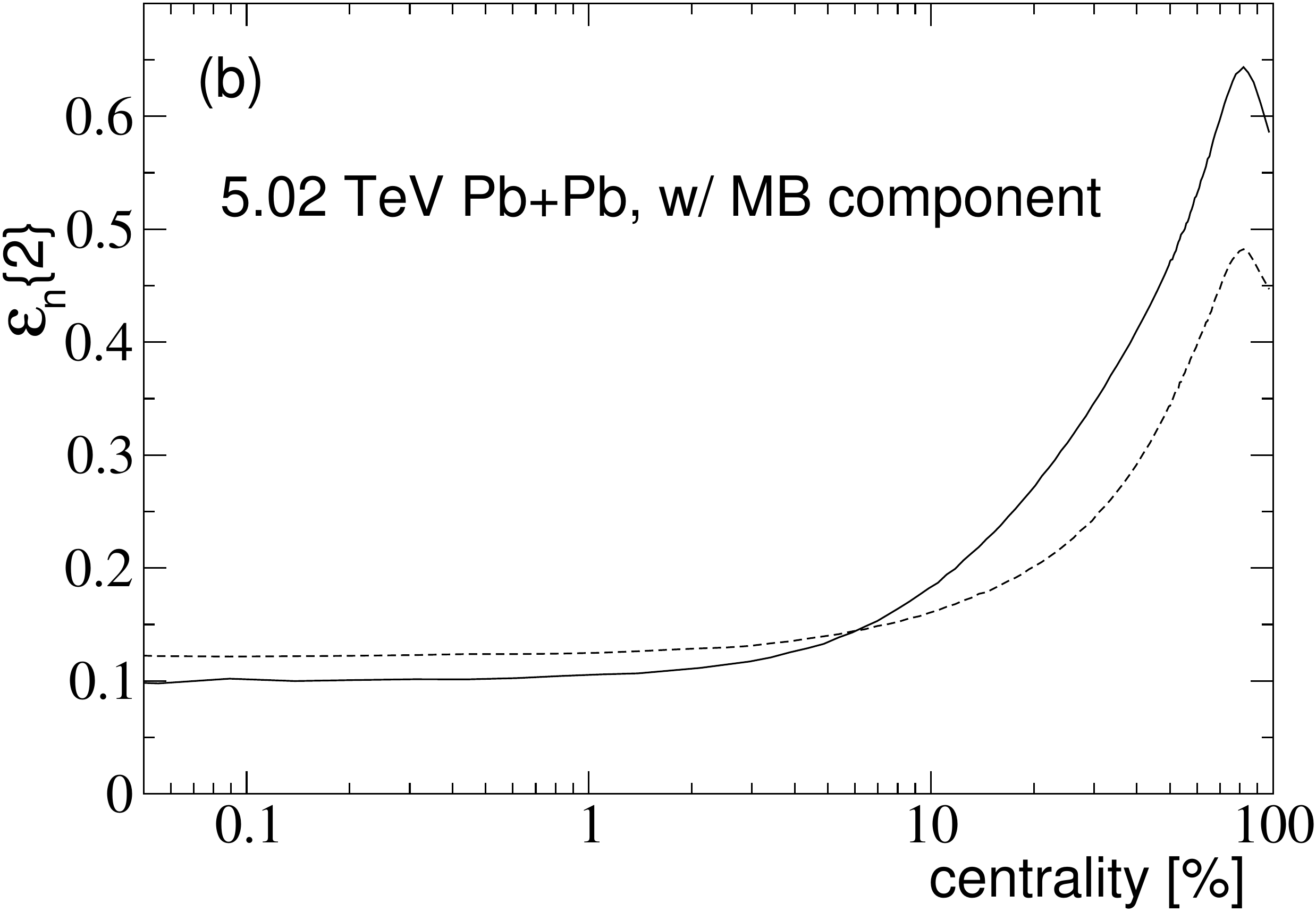}  
\end{center}
\caption[.]{The rms azimuthal coefficients $\epsilon_{2}\{2\}$ (solid)
  and $\epsilon_3\{2\}$ (dashed) vs centrality
  for 5.02 TeV Pb+Pb collisions obtained within the MC Glauber model
  without (a) and with (b) the meson-baryon component of the nucleon
  using the MC WS sampling of the nucleon positions with
  the smooth $Q_2^2$-filter that gives
$\langle Q_2^2 \rangle=\langle Q_2^2 \rangle_{WS}/r_2$
with  $r_2=2.25$ (see text for explanations). }
\end{figure}
\begin{figure}[!h] 
\begin{center}
\includegraphics[height=5.5cm]{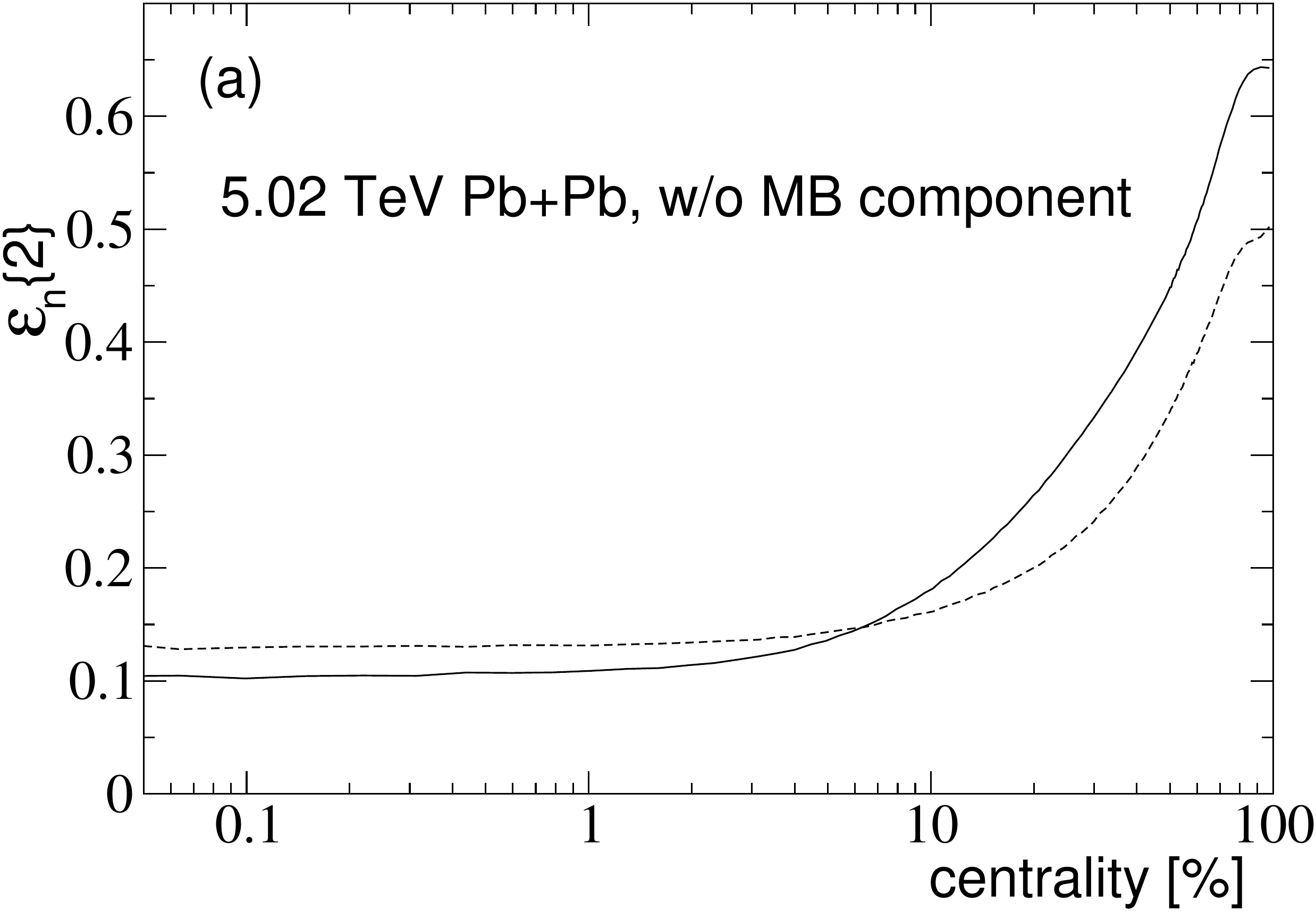}  
\includegraphics[height=5.5cm]{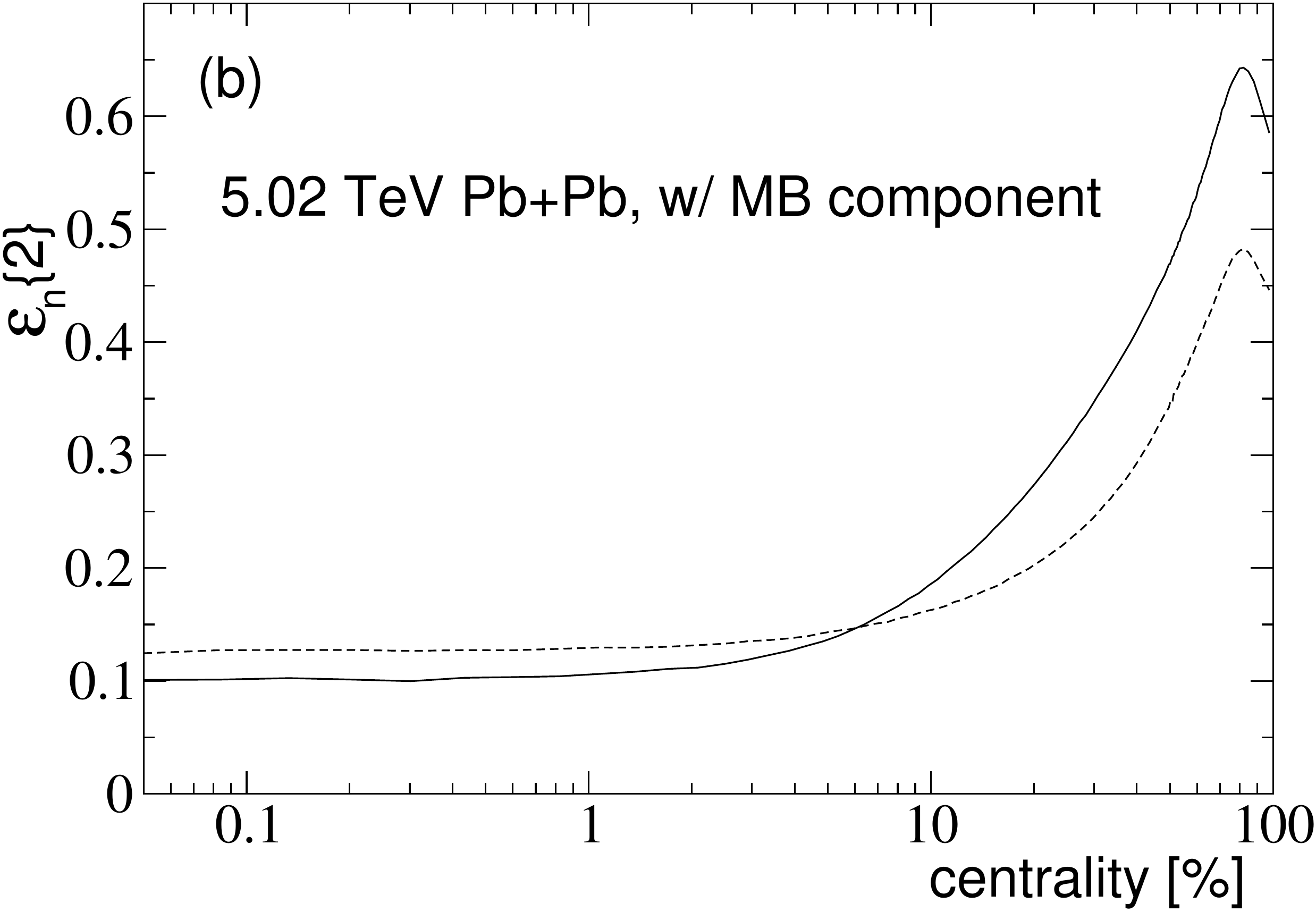}  
\end{center}
\caption[.]{The rms azimuthal coefficients $\epsilon_{2}\{2\}$ (solid)
  and $\epsilon_3\{2\}$ (dashed) vs centrality
  for 5.02 TeV Pb+Pb collisions obtained within the MC Glauber model
  without (a) and with (b) the meson-baryon component of the nucleon
  using the MC WS sampling of the nucleon positions with
  the smooth $Q_2^2$-
and $Q_2^2$-filters
  that give
$\langle Q_2^2 \rangle=\langle Q_2^2 \rangle_{WS}/r_2$
  with  $r_2=2.25$
and $\langle Q_3^2 \rangle=\langle Q_3^2 \rangle_{WS}/r_3$
  with  $r_3=0.84$
  (see text for explanations). }
\end{figure}
\begin{figure}[!h] 
\begin{center}
\includegraphics[height=5.5cm]{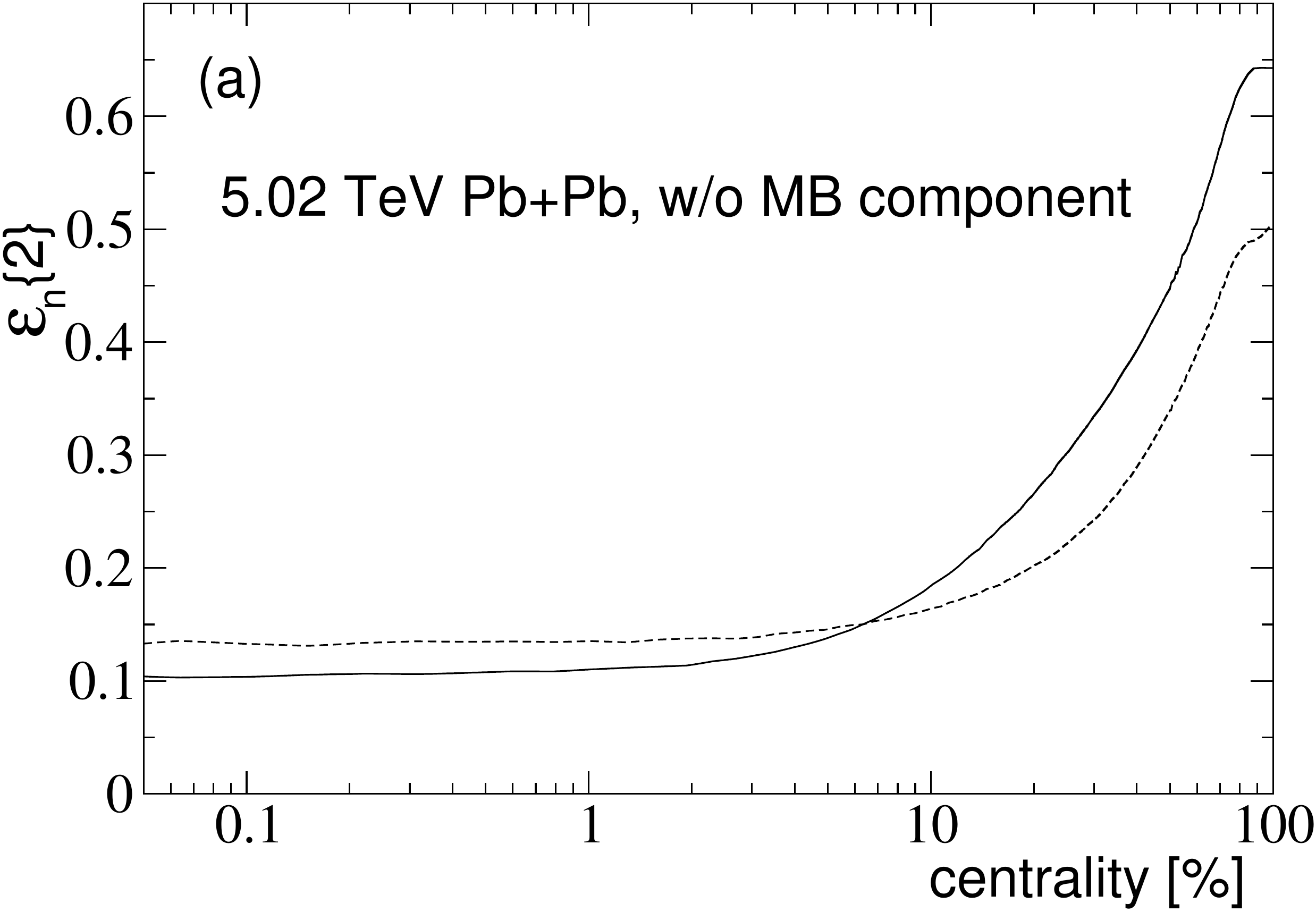}  
\includegraphics[height=5.5cm]{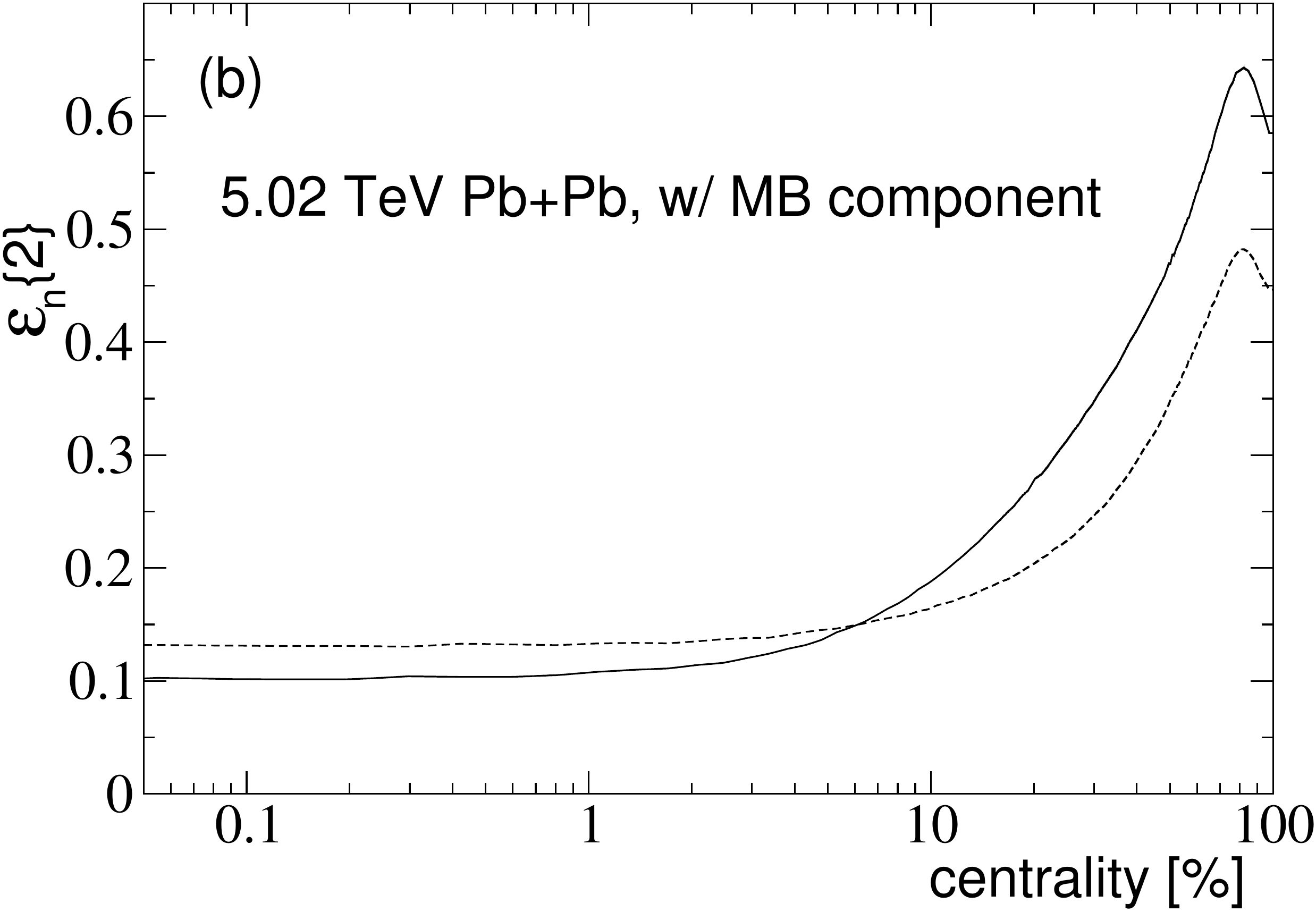}  
\end{center}
\caption[.]{Same as in Fig. 4 for
$r_3=0.7$. }
\end{figure}

In this section we present our numerical results for the rms ellipticity
$\epsilon_2\{2\}$ and
triangularity $\epsilon_3\{2\}$ for 5.02 TeV Pb+Pb collisions \footnote{
  Note that our calculations show that the $Q_{2}^2$- and $Q_3^2$-filterings
  give  almost zero effect on the higher
harmonics $\epsilon_4$ and $\epsilon_5$, and hence we do not show them.}.
The results for
2.76 TeV Pb+Pb collisions are very close to those for 5.02 TeV, and
hence we do not show them.
For the versions with the $Q_{2,3}^2$-filtering, we present the results
obtained with the smooth filters (as we already said, the 
results for the versions with
the smooth and sharp $Q_{2,3}^2$-filters are practically indistinguishable).
The results have
been obtained by generating $\sim 6\cdot 10^6$ Pb+Pb collisions, i.e.
we have about $6\cdot 10^4$ events in the region $c\lsim 1$\%, which
is most interesting in the context of the $v_2$-to-$v_3$ puzzle.
We have performed calculations for the Glauber schemes
with and without the meson-baryon component of the nucleon.
We present the results obtained for the entropy sources with
the Gaussian width parameter $\sigma=0.4$ fm. In the small
centrality region ($\lsim 5-10$\%), that is interesting in the context
of the $v_2$-to-$v_3$ puzzle, the predictions for $\epsilon_{2,3}\{2\}$
have very low sensitivity to the value of $\sigma$.
We have checked this by performing calculations for the
value $\sigma=0.7$ fm. In this case, $\epsilon_{2,3}\{2\}$
become somewhat smaller at large centralities (by $\sim 5-7$\%
at $c\sim 50$\%), but at centrality $\lsim 5-10$\% the results
are very close to those for $\sigma=0.4$ fm.

In Figs. 2a and 2b we show the results for centrality dependence
of $\epsilon_{2,3}\{2\}$ for the ordinary MC WS sampling of the nucleon
positions (i.e., without applying any $Q_{2,3}^2$-filters).
From Figs. 2a and 2b one can see that the results for $\epsilon_{2,3}\{2\}$
in the  versions without and with the meson-baryon component are very similar.
For the curves shown in Fig. 2,  we have on average 
$\epsilon_2\{2\}/\epsilon_3\{2\}\approx 0.94-0.95$ at $c\lsim 1$\%.
In Figs. 3a and 3b  we show  $\epsilon_{2,3}\{2\}$
obtained with the MC sampling of the nucleon
positions with applying the smooth $Q_2^2$-filter, which gives for
the colliding nuclei $\langle Q_2^2 \rangle=\langle Q_2^2 \rangle_{WS}/r_2$
  with  $r_2=2.25$, i.e.,
  the mean squared quadrupole moment consistent with the EWSR prediction.
  From comparison of the results shown in Figs. 2 and 3, one case see
  that the presence of the $Q_2^2$-filter
  noticeably reduces $\epsilon_2\{2\}$, but almost does not affect
  $\epsilon_3\{2\}$.
For Fig. 3, at $c\lsim 1$\% we have on average
$\epsilon_2\{2\}/\epsilon_3\{2\}\approx 0.82-0.84$.
Note that the value of the ratio $\epsilon_2\{2\}/\epsilon_3\{2\}$
at $c\lsim 0.1$\% for the curves shown in Fig.~3, is just by $\sim 2$\%
larger than that obtained in Ref. \cite{Z1} in similar calculations
for zero impact parameter. 

In Figs. 4a and 4b we plot $\epsilon_{2,3}\{2\}$ obtained
  with the MC sampling of the nucleon positions
  with applying simultaneously the smooth $Q_2^2$- and $Q_3^2$-filters,
that give for
the colliding nuclei
$\langle Q_{2,3}^2 \rangle=\langle Q_{2,3}^2 \rangle_{WS}/r_{2,3}$
with  $r_2=2.25$  and $r_3=0.84$.
The addition of the $Q_3^2$-filtering for $r_3=0.84$ increases
$\epsilon_3\{2\}$
by $\sim 2$\% at $c\lsim 1$\%, and in this centrality region
we have now on average
$\epsilon_2\{2\}/\epsilon_3\{2\}\approx 0.8-0.82$.
In Fig. 5 we show the results similar to that plotted in Fig. 4,
but for $r_3=0.7$. In this version, at $c\lsim 1$\% we have on average
$\epsilon_2\{2\}/\epsilon_3\{2\}\approx 0.78-0.81$.
From comparison of the results shown in Fig. 3 with that shown in Figs. 4
and 5, one can see that the $Q_3^2$-filtering increases slightly
$\epsilon_3\{2\}$, without a noticeable effect on the value of
$\epsilon_2\{2\}$.
The results shown in Figs. 3, 4  and 5 demonstrate that the effect of the
$Q_{2,3}^2$-filters
becomes seen only at $c\lsim 10$\%.
From Figs. 3, 4  and 5, one can see that,
in the most interesting (in the
context of the $v_2$-to-$v_3$ puzzle) region
of small centralities $c\lsim 1$\%,
modification of the MC sampling of the nucleon positions
with the $Q_2^2$- and $Q_3^2$-filters increases the difference
$\epsilon_3\{2\}-\epsilon_2\{2\}$ by a factor of $\sim 3$.
Note that our values for the ratio 
$\epsilon_2\{2\}/\epsilon_3\{2\}$ at $c\lsim 1$\%, for the versions with
the $Q_{2,3}^2$-filtering,
are smaller than those obtained within the MC-KLN model in
Ref. \cite{v23_Heinz}
 by $15-20$\%, and by $\sim 10-15$\%
than obtained within the TRENTO scheme in Ref. \cite{v23_Luzum}
(for the octupole deformation parameter $\beta_3\sim 0-0.0375$).
 As compared to calculations of Ref. \cite{Snigirev} within the MAGMA model
 our values of the ratio $\epsilon_2\{2\}/\epsilon_3\{2\}$ are smaller by a
 factor of $\sim 1.65$. 

From the point of view the $v_2$-to-$v_3$ puzzle, it is interesting
to know the ratio $k_2 \epsilon_2\{2\}/k_3\epsilon_3\{2\}$.
Hydrodynamic simulations of Pb+Pb collisions at the LHC energies
give $k_2/k_3\approx 1.2-1.4$
\cite{Olli_k23,v23_Heinz,Olli_Xe,v23_Luzum}
for small centralities ($c\lsim 2$\%).
Our results shown in Figs. 4 and 5
with the MC sampling of the nucleon
positions with the simultaneous $Q_{2}^2$-filtering (with $r_2=2.25$)
and $Q_3^2$-filtering
for centrality $\sim 0.1-0.2$\% give
$\epsilon_2\{2\}/\epsilon_3\{2\}\approx 0.8(0.78)$ at
$r_3=0.84(0.7)$.
These values of $\epsilon_2\{2\}/\epsilon_3\{2\}$ lead to
$0.96(0.94)<k_2 \epsilon_2\{2\}/k_3\epsilon_3\{2\}< 1.12(1.1)$
for $r_3=0.84(0.7)$ and $1.2<k_2/k_3<1.4$.
This agrees reasonably
with the ALICE measurements \cite{ALICE_v23}
for 2.76 and 5.02 TeV Pb+Pb collisions that
give at $c\to 0$ $v_2\{2\}/v_3\{2\}\approx 1.08\pm 0.05$.

  The above results have been obtained
  for the uncorrelated WS nuclear density.
We also performed calculations replacing it
by the WS nuclear density with the hard $NN$ repulsion for
the expulsion radius $r_c=0.9$ \cite{Bron_NNcore} and $0.6$ \cite{HC_HRG} fm.
We have found that
the $NN$ hard core changes slightly the values of
$\epsilon_{2,3}$ for MC simulations without the $Q_{2,3}^2$-filtering.
However, for the version with the simultaneous $Q_{2,3}^2$-filtering
predictions for $\epsilon_{2,3}\{2\}$
are very close to those for the uncorrelated WS nuclear density.
This fact shows that predictions for $\epsilon_{2,3}\{2\}$
depend mostly on the large-scale ($L\sim R_A$) 
properties of the many-body nuclear distribution, and its properties 
on the small-scale distances ($L\sim r_c\ll R_A$)  are of minor
importance.
This may be viewed as another argument in favor of our basic idea to
model the collective effects in the $^{208}$Pb nucleus by applying suitable
$Q_{2,3}^2$-filterings
of the nucleon positions in the MC simulations, which
guarantee that the selected set of the nucleon positions reproduces
the EWSR predictions for
$\langle Q_{2,3}^2\rangle$.

In connection with modeling the effect of the $NN$ hard core
in MC simulations of $AA$ collisions,
it worth noting that it is not evident that
models with the excluded volume
are  physically better justified
than simulations with the uncorrelated WS nuclear density.
The point is that it is possible that in reality
the ``excluded volume'' is not empty.
Indeed, 
the short range $NN$ interaction
can be successfully described in the dibaryon picture
(for reviews, see \cite{dibaryon1,Simonov2}), in which
the expulsion region is not empty, but occupied by a $6q$-cluster.
In this case, similarly to $hD$-scattering \cite{Z_6q}, the $6q$-clusters
should participate in the $t$-channel gluon exchanges
between the colliding nuclei and contribute to the entropy production
in $AA$ collisions. It is clear that in this scenario the use of
the uncorrelated WS nuclear density is more adequate for simulation
of the initial conditions in heavy ion collisions.

\section{Conclusions}
The present study is an extension of our previous analysis
\cite{Z1} of the influence of the collective quantum
effects in the nuclear many-body distribution on the anisotropy
coefficients $\epsilon_{2,3}$ in Pb+Pb collisions at the LHC
energies, motivated by the $v_2$-to-$v_3$ puzzle in ultra-central
Pb+Pb collisions. Contrary to our previous calculations \cite{Z1}, where
only collisions at a zero impact parameter have been studied,
we perform calculations for the $n$-centrality and in
the whole centrality range.
We model the collective effects in the colliding Pb nuclei by modifying
the MC sampling of the nucleon positions by using suitable filters
that guarantee
that the mean squared quadrupole and octupole moments
coincide with the ones
obtained using the EWSR from the data on the quadrupole and octupole
strength functions of the $^{208}$Pb nucleus.
We have found that the EWSR and experimental data on
the ISGQR of the $^{208}$Pb nucleus lead to
the mean squared quadrupole moment that is smaller than the one for
the uncorrelated WS nuclear
density by the factor $r_2\approx 2.25$. For the octupole mode,
the available experimental data on the octupole strength function
support that the ratio between the mean squared octupole moment
for the uncorrelated WS nuclear density
and the one obtained with the EWSR should be $\sim 0.7-0.84$.

We have performed the MC Glauber model calculations with applying
the smooth and sharp
$Q_{2,3}^2$-filters to generate the sample of the nucleon positions.
We find that the results for $\epsilon_{2,3}\{2\}$
obtained with the smooth and sharp $Q_{2,3}^2$-filterings are practically
identical.
Our numerical results show that the effect of the $Q_{2,3}^2$-filtering
of the nucleon positions on the values of
$\epsilon_{2,3}\{2\}$ becomes seen at $c\lsim 10$\%.
At centralities $c\sim 0.1-1$\% our MC Glauber model
with the modified sampling of the nucleon positions
gives to $\epsilon_2\{2\}/\epsilon_3\{2\}\sim 0.8$, which is by a factor of
$\sim 1.2$
smaller than that for the ordinary MC sampling of the nuclear
positions for the uncorrelated
WS nuclear density. 
Such a value of the ratio $\epsilon_2\{2\}/\epsilon_3\{2\}$
allows to reach a reasonable agreement with the ratio
$v_2\{2\}/v_3\{2\}\approx 1.08\pm 0.05$ at $c\to 0$
obtained for 2.76 and 5.02 TeV Pb+Pb collisions
by ALICE \cite{ALICE_v23} for the ratio $k_2/k_3\approx 1.35$, 
which is consistent with the window $1.2<k_2/k_3<1.4$
supported by the hydrodynamic simulations
of Refs. \cite{Olli_k23,v23_Heinz,Olli_Xe,v23_Luzum}.

Although, our analysis demonstrates the importance of the collective
effects for the geometry of the initial QGP fireball
for a spherical nucleus, one can expect that the collective effects
may be important for collisions of the non-spherical nuclei as well (e.g. for
$^{197}$Au+$^{197}$Au and $^{238}$U+$^{238}$U collisions).
The collective effects may be important for nuclear shape investigation
 \cite{Jia1} and
for interpretation of the results of
the event shape engineering \cite{shape_eng_Vol,STAR_UU,shape_eng_Heinz}
in $AA$ collisions at the RHIC and LHC energies,
and at the NICA energy region, where
the critical point effects may affect the medium expansion,
and the account of suppression of the quadrupole fluctuations
for the Au nucleus is especially important. \\

\begin{acknowledgments}
  I am grateful to S.P. Kamerdzhiev  for helpful discussions on
physics of the giant resonances and our method of calculation of the
squared $L$-multipole moments.
This work was partly supported by the RFBR grant 
18-02-40069mega.
\end{acknowledgments}

\begin{appendix}
\renewcommand{\theequation}{A\arabic{equation}}
\setcounter{equation}{0}

\section*{Appendix: Calculation of the
  mean squared multipole moments of the $^{208}$Pb nucleus}

For completeness, in this appendix we briefly review the method
of Ref.~\cite{Z1} for calculation of the mean squared
multipole moments of the $^{208}$Pb nucleus
with the help of the EWSR \cite{BM,EWSR},
and give the ratios between the mean squared multipole moments
obtained using the ordinary MC WS sampling of the nucleon positions
and those calculated using the EWSR.

We assume that  in the ground state the $^{208}$Pb nucleus 
is spherical. We write the nuclear density in  the WS form
\beq
\rho_{A}(r)=\frac{\rho_0}{1+\exp[(r-R_A)/d]}\,
\label{eq:a10}
\eeq
with $R_{A}=6.62$ fm, and $d=0.546$ fm \cite{PHOBOS,ATDATA}.
We define the quadrupole and octupole moments in terms
of the spherical harmonics, $Y_{Lm}$, with $L=2$ and $3$.
The needed isosinglet $L$-multipole operator
reads (see, e.g. \cite{BM,Greiner,Roca})
\beq
F_L=\sum_{i=1}^A r^L_iY_{Lm}(\nb_i)
\label{eq:a20}
 \eeq
 with ${\nb}_i=\rb_i/|\rb_i|$.
 The mean squared $L$-multipole moment, $\langle Q_L^2\rangle$,
of a nucleus in the ground state can be defined
quantum mechanically as
\beq
\langle Q_L^2\rangle=
\langle 0 |F^{+}_LF_L|0\rangle\,.
\label{eq:a30}
\eeq

Classical calculation of $\langle Q_L^2\rangle$ for the uncorrelated
WS nuclear density gives\footnote{We ignore in this appendix
  a very small effect of the c.m. nucleon correlations. However,
  in our numerical simulations they have been treated properly.}
  \beq
\langle Q_L^2\rangle_{WS}=\langle F^{+}_{L}F_L\rangle_{WS}=\frac{A(2L+1)\langle
    r^{2L}\rangle}{4\pi}\,.
  \label{eq:a40}
  \eeq
Of course, this formula becomes invalid if one includes the effect
of the short range hard core $NN$ correlations. But their effect
is not very strong (see below).
To perform quantum calculation of $\langle Q_L^2\rangle$
of the $^{208}$Pb nucleus
we use the EWSR (for a review, see \cite{EWSR})
for strength function, $S(\omega)$ of the operator $F_L$.
It is defined as
\beq
S(\omega)=\sum_n |\langle n|F_L|0\rangle |^2\delta(\omega-\omega_n)\,,
\label{eq:a50}
  \eeq
  where $\omega_n=E_n-E_0$ and $E_n$ are the nucleus state energies.
  In terms of moments of the strength function, given by
  \beq
  m_k= \int_0^{\infty} d\omega \omega^k S(\omega)\,,
  \label{eq:a60}
    \eeq
we can write $\langle 0 |F^{+}_LF_L|0\rangle=m_0$.
    It is convenient to rewrite it as
\beq    
\langle 0 |F^{+}_LF_L|0\rangle=\frac{m_1}{E_c}\,,
\label{eq:a70}
 \eeq
where
  \beq
  E_c=m_1/m_0
  \label{eq:a80}
  \eeq
is the so called the centroid energy $E_c$, 
which can be viewed as the typical excitation energy
for the operator $F_L$ acting on the ground state.
The representation (\ref{eq:a70}) is more convenient than the
one via $m_0$, because the experimental errors in the
normalization of the strength function are not important
for the ratio $m_1/m_0$, and the moment
$m_1$ can be exactly calculated with the help of
the EWSR, which for $L\ge 2$ \cite{BM,EWSR,Roca}
gives 
 \beq
 m_1=\frac{AL(2L+1)^2\langle r^{2L-2}\rangle}{8\pi m_N}\,,
 \label{eq:a90}
 \eeq
 where $m_N$ is the nucleon mass.
 Thus, we have
\beq
 \langle Q_L^2\rangle_{EWSR}=
 \frac{AL(2L+1)^2\langle r^{2L-2}\rangle}{8\pi m_NE_c}\,.
\label{eq:a100}
 \eeq 
 Comparing (\ref{eq:a100}) with (\ref{eq:a40}), we see that
 the ratio between the mean squared multipole moments
for the ordinary MC sampling of the nucleon positions and that
for quantum calculation with the help of the EWSR reads 
\beq
r_L=\frac{\langle Q_L^2\rangle_{WS}}
      {\langle Q_L^2\rangle_{EWSR}}=
\frac{ 2m_NE_c \langle r^{2L}\rangle}
     {L(2L+1)\langle r^{2L-2}\rangle}\,.
     \label{eq:a110}
      \eeq

We calculate
the centroid energy using the Breit-Wigner parametrization of
the strength function. Since the strength function is proportional to
the imaginary part of the polarisability(susceptibility) $\alpha$
(which, as usual, should  satisfy the relation
$\alpha(-\omega^*)=\alpha^*(\omega)$ \cite{LL5})
for the operator $F_L$, for each resonance
a double Breit-Wigner parametrization
with the poles at $\pm\omega_R-i\Gamma_R/2$
(with the same
residues)
should be used  (see Eq. (20) of \cite{Z_e2}).
For $N$ resonances this gives
\beq
E_c=\left[\sum_{i=1}^{N}\frac{2f_i}{\pi\omega_{i}}\mbox{arctg}{2\omega_i/\Gamma_i}
  \right]^{-1}\,
\label{eq:a120}
\eeq  
with $f_i$ the fraction of the $i$th resonance contribution to the EWSR.

For the isoscalar $F_2$ operator, the EWSR for the $^{208}$Pb nucleus 
is practically exhausted by the isoscalar giant quadrupole
resonance  with $\omega\approx 10.89$ MeV and
$\Gamma\approx 3$ MeV \cite{Youngblood_PRC69}. Formula (\ref{eq:a120}) with
these parameters 
gives
$E_c\approx 11.9$ MeV,
then from (\ref{eq:a110}) one can obtain $r_2\approx 2.25$.
Thus, we see that probabilistic treatment
of the $^{208}$Pb nucleus with the WS nuclear density
overestimates the 3D quadrupole fluctuations.
It is clear that this can lead to incorrect predictions
for the 2D fluctuations of the initial QGP fireball
in $AA$ collisions as well. As in \cite{Z1},  our strategy to cure
this problem is to modify the MC sampling of the nucleon positions
by applying a suitable filter that generates the nuclear configurations
with the mean squared quadrupole moment consistent with the EWSR.

To calculate $r_3$ we need the strength function for $F_3$.
For the $^{208}$ Pb nucleus, the function $S(\omega)$ for the operator $F_3$
is distributed in a
broad range of $\omega$. 
There are several very narrow peaks in the low-energy region 
$\omega\lsim 7$ MeV,
\cite{Harakeh_NPA327,Yamagata_PRC23,Fujita_LEOR1992},
in which
the low lying $3^-$ state with $\omega\approx 2.615$ MeV
exhausts $\sim 20-25$\%
of the EWSR \cite{Yamagata_PRC23,Harakeh_NPA327,Fujita_LEOR1992}
and several more states in the region
$ 4.7 \lsim \omega \lsim 7$ MeV (so called the low-energy octupole resonance
(LEOR) region) that exhaust about 
$8-13$\% of the EWSR \cite{Harakeh_NPA327,Fujita_LEOR1992}.
In the high-energy region there is a broad resonance at 
$\omega\sim 16-20$ MeV with $\Gamma\sim 5-8$ Mev
\cite{Yamagata_PRC23,Pitthan_HEOR,Sasao,Carey_HEOR,Davis_HEOR,Youngblood_PRC69}.
The measured EWSR fraction of the high-energy octupole resonance (HEOR)
varies from $\sim 20-50$\% \cite{Carey_HEOR,Sasao}
to $\sim 60-90$\%
\cite{Pitthan_HEOR,Davis_HEOR,Yamagata_PRC23,Youngblood_PRC69}.
Using the data from Ref. \cite{Fujita_LEOR1992}, that give
$21$\% for the EWSR fraction of the $2.615$ MeV $3^-$ state,
and $8.3$\% for the EWSR fraction of the LEOR region, together
with parameters of the HEOR from Ref. \cite{Youngblood_PRC69}
($\omega\approx 19.6\pm 0.5$ MeV, $\Gamma\approx 7.4\pm 0.6$ MeV with
the EWSR fraction $70\pm14$\%) we obtain
$r_3\approx 0.84$. However, if we take $25$\% for the EWSR fraction
of the $2.615$ MeV state as obtained in \cite{Yamagata_PRC23},
and the parameters of the HEOR obtained in \cite{Sasao}
($\omega=16$ MeV, $\Gamma=6$ MeV),
then we obtain $r_3\approx 0.7$. Thus,
we see that the experimental data on the octupole strength
function of the $^{208}$Pb nucleus support that $r_3\lsim 1$. 
But due to the experimental uncertainties for the octupole strength
function, we have uncertainties in the value of $r_3$
about $15-20$\%. In the present analysis we perform
calculations for two values $r_3=0.84$ and $0.7$.

The above values of the coefficients $r_2$ and $r_3$ correspond to
the MC sampling of the nucleon positions with the uncorrelated WS
nuclear density. Calculations using the WS distribution with
restrictions on the minimum nucleon-nucleon distances,
to mimic the $NN$ hard core,
give somewhat different values of $r_{2,3}$. 
However, the effect of the $NN$ hard core on $r_{2,3}$
is relatively small:  we obtained the reduction of $r_2$
by the factor $0.78(0.926)$, and the reduction
of $r_3$ by the factor $0.81(0.928)$ for
the core radius $r_c=0.9(0.6)$ fm.

In the present analysis we modify the MC sampling of the nucleon
positions only with the filters for the isoscalar $L=2$ and $3$
moments, which correspond to
the nuclear shape fluctuations. We do not use a filter for the
$L=0$ mode, that corresponds to the pure radial fluctuations.
The radial fluctuations may be characterized by the squared moment for
the monopole isoscalar operator
$F_0=\sum_{i=1}^{A}(r_i^2-\langle r^2\rangle)$.
The EWSR for this operator gives $m_1=2\langle r^2\rangle/m_N$
\cite{compres1}. Using this formula, 
for the uncorrelated WS nuclear density, we obtain
for the analogue of (\ref{eq:a110}) in the case of the $L=0$ mode
\beq
r_0=\frac{m_N E_c}{2}\left[\frac{\langle r^4\rangle}
  {\langle r^2\rangle}-\langle r^2\rangle\right]\,.
\label{eq:a130}
\eeq
For the isoscalar $L=0$ mode the EWSR is practically exhausted by
the isoscalar giant monopole resonance with $\omega\approx 13.6-13.9$
MeV and $\Gamma\approx 3$ MeV \cite{Youngblood_PRC69,Patel_2014}.
These parameters give $E_c\approx 15$ MeV, and calculation using
(\ref{eq:a130}) for the WS distribution (\ref{eq:a10}) gives $r_0\sim 1.6$.
This means that
for the MC sampling of the nuclear configurations with the uncorrelated WS
nuclear density the magnitude of the pure radial fluctuations
is somewhat overpredicted as compared to that extracted from the experimental monopole strength
function. However, we have found that adding the filtering for the $L=0$
mode, that decreases  the mean squared $L=0$ moment to its EWSR value,
practically does not affect the azimuthal coefficients $\epsilon_{2,3}$.
Therefore we do not use filtering for the $L=0$ fluctuations.

\end{appendix}

\end{document}